\documentclass[showpacs,preprintnumbers]{revtex4}
\usepackage{amsmath}
\usepackage{graphicx}

\begin{document}

\textbf{Comment on \textquotedblleft Limit on the Electron Electric Dipole
Moment in Gadolinium-Iron Garnet\textquotedblright }\bigskip

Tomislav Ivezi\'{c}

\textit{Ru%
\mbox
{\it{d}\hspace{-.15em}\rule[1.25ex]{.2em}{.04ex}\hspace{-.05em}}er Bo\v{s}%
kovi\'{c} Institute, P.O.B. 180, 10002 Zagreb, Croatia}

\textit{ivezic@irb.hr\bigskip }

\noindent PACS number(s): 32.10.Dk, 11.30.Er, 14.60.Cd, 75.80.+q\bigskip

In [1], a solid-state electron electric dipole moment (EDM) experiment is
described in which a sample (GdIG) electric polarization appears when the
sample is magnetized. The voltage is induced across a GdIG sample by the
alignment of the sample's magnetic dipole moments (MDMs) in an applied
magnetic field $\mathbf{H}$. All solid-state electron EDM experiments rely
on the fact that the electron EDM $\mathbf{d}$ is collinear with its MDM $%
\mathbf{m}$, because they are both supposed to be proportional to the spin $%
\mathbf{S}$; it is supposed that $\mathbf{S}$ is the only available 3-vector
in the rest frame of the particle. Thus, the interaction of the EDMs and $%
\mathbf{H}$ is only indirect through the alignment of MDMs, i.e., the
three-dimensional (3D) spins, in the field $\mathbf{H}$.

Recently, [2], the Uhlenbeck-Goudsmit hypothesis is generalized in a Lorentz
covariant manner using 4D geometric quantities; the dipole moment tensor $%
D^{ab}$ is proportional to the spin four-tensor $S^{ab}$, $%
D^{ab}=g_{S}S^{ab} $, Eq. (9) in [2]. The dipole moment vectors $d^{a}$ and $%
m^{a}$ are derived from $D^{ab}$ and the velocity vector of the particle $%
u^{a}$, Eq. (2) in [2]. Similarly, the usual 4D spin $S^{a}$, and a new one,
the intrinsic angular momentum, spin $Z^{a}$, are derived from $S^{ab}$ and $%
u^{a}$, Eq. (8) in [2]. Then, Eq. (10) in [2] is obtained as $%
m^{a}=cg_{S}S^{a}$,$\ d^{a}=g_{S}Z^{a}$. Accordingly, the intrinsic MDM $%
m^{a}$ is determined by $S^{a}$, whereas the intrinsic EDM $d^{a}$ is
determined by the new spin vector $Z^{a}$ and not, as usual, by the spin $%
\mathbf{S}$. Both spins, $S^{a}$ and $Z^{a}$, are equally good physical
quantities.

Instead of an indirect interaction between the applied $\mathbf{H}$ and an
EDM $\mathbf{d}$ through the alignment of 3D spins by the interaction $-m(%
\mathbf{S}/S)\cdot \mathbf{B}$, we propose a direct, Lorentz covariant,
interaction between $B^{a}$ and an EDM $d^{a}$.

Inserting the decomposition of $F^{ab}$ (in terms of $E^{a}$, $B^{a}$ and
the velocity vector $v^{a}$ of the observers who measure fields), Eq. (1) in
[2], and that one of $D^{ab}$, Eq. (2) in [2], into the interaction term $%
(1/2)F_{ab}D^{ba}$, one finds Eq. (3) in [2] (it is first reported in [3]).
When it is taken that the laboratory frame is the $e_{0}$-frame (the frame
in which the observers who measure $E^{a}$ and $B^{a}$ are at rest, with the
standard basis $\{e_{\mu }\}$ in it), then $E^{0}=B^{0}=0$, and \emph{only}
three spatial components $E^{i}$ and $B^{i}$ will remain. Similarly, \emph{%
only} in the particle's rest frame, with the standard basis in it, $%
d^{0}=m^{0}=0$ and only $d^{i}$ and $m^{i}$ will remain. Hence, it is not
possible that, e.g., in the laboratory frame, \emph{both}, the fields and
the dipole moments have \emph{only} three spatial components, i.e., as for
the usual 3-vectors. (In all EDM experiments the interaction is described in
terms of the 3-vectors as $\mathbf{E}\cdot \mathbf{d}$ and $\mathbf{B}\cdot
\mathbf{m}$.)

In the laboratory frame as the $e_{0}$-frame, one can neglect the
contributions to $L_{int}$, Eq. (3) in [2], from the terms with $d^{0}$ and $%
m^{0}$; they are $u^{2}/c^{2}$ of the usual terms $E_{i}d^{i}$ or $%
B_{i}m^{i} $. Then, what remains is
\begin{equation}
L_{int}=-((E_{i}d^{i})+(B_{i}m^{i}))-(1/c^{2})\varepsilon
^{0ijk}(E_{i}m_{k}-c^{2}B_{i}d_{k})u_{j}.  \label{1}
\end{equation}%
With the usual 3-vectors, it would correspond to Eq. (26) in [2]. But, as
stated in [2]: \textquotedblleft ... what is essential for the number of
components of a vector field is the number of variables on which that vector
field depends, i.e., the dimension of its domain. Thus, strictly speaking,
the time-dependent $\mathbf{E(r,}t\mathbf{)}$ and $\mathbf{B(r,}t\mathbf{)}$
cannot be the 3-vectors, since they are defined on the
spacetime.\textquotedblright\ Furthermore, as noticed in [2]:
\textquotedblleft ... neither the direction of $\mathbf{d}$ nor the
direction of the spin $\mathbf{S}$ have a well-defined meaning in the 4D
spacetime. The only Lorentz-invariant condition on the directions of $d^{a}$
and $S^{a}$ in the 4D spacetime is $d^{a}u_{a}=S^{a}u_{a}=0$. This condition
does not say that $\mathbf{d}$ has to be parallel to the spin $\mathbf{S}$%
.\textquotedblright\ Obviously, the same remark holds if $\mathbf{d}$ is
replaced by $\mathbf{m}$ and $d^{a}$ by $m^{a}$. The results from [2]
indicate that the basic points of the interpretation of measurements of EDM
in [1], i.e., both $\mathbf{m}$ \emph{and} $\mathbf{d}$ are parallel to $%
\mathbf{S}$, are meaningless in the manifestly covariant formulation from
[2]. This means that the usual formulation with 3-vectors $\mathbf{E}$, $%
\mathbf{B}$, $\mathbf{S}$, etc. \textbf{IS NOT }relativistically correct
formulation.

It is seen from Eq. (\ref{1}) that the interaction between $B^{a}$ and a MDM
$m^{a}$ is contained in the term $-B_{i}m^{i}$.and that one between $B^{a}$
and an EDM $d^{a}$ is contained in the term $\varepsilon
^{0ijk}B_{i}d_{k}u_{j}$, which is $u^{a}$ - dependent.

In conclusion, according to Eq. (\ref{1}), a voltage induced across the
solid is not caused by the alignment of MDMs in an applied field $\mathbf{H}$
than by the polarization of the sample due to the interaction $\varepsilon
^{0ijk}B_{i}d_{k}u_{j}$. That voltage can give some information about $d^{a}$%
, because that term contains the EDM $d^{a}$. \bigskip

\noindent \textbf{References\bigskip }

\noindent \lbrack 1] B. J. Heidenreich \textit{et al}., .Phys. Rev. Lett.
\textbf{95}, 253004 (2005).

\noindent \lbrack 2] T. Ivezi\'{c}, Phys. Scr. \textbf{81}, 025001 (2010).

\noindent \lbrack 3] T. Ivezi\'{c}, Phys. Rev. Lett. \textbf{98}, 108901
(2007).

\end{document}